\documentstyle[prb,aps,multicol]{revtex}
  \renewcommand{\narrowtext}{\begin{multicols}{2} \global\columnwidth20.5pc}
  \renewcommand{\widetext}{\end{multicols} \global\columnwidth42.5pc}

\input{psfig}

\newcommand{\f}[1]{Fig.~\ref{#1}}
\newcommand{\eq}[1]{Eq.~(\ref{#1})}
\newcommand{\eqs}[2]{Eqs.~(\ref{#1}) and~(\ref{#2})}

\def\be{\begin{equation}}
\def\ee{\end{equation}}
\def\bea{\begin{eqnarray}}
\def\eea{\end{eqnarray}}
\def\l({\left(}
\def\r){\right)}
\def\p{\protect}
\def\Bn{B_0}
\def\Bam{B_{am}}
\def\Jc2{\tilde J_c}
\def\Bceff{B_c^{\text{eff}}}
\def\Mrev{M_{\rm rev}}

\begin{document}
\bibliographystyle{prsty}

\title {Thin superconducting disk with
field-dependent critical current: \\
Magnetization and ac susceptibilities}
\author{D.~V. Shantsev$^{1,2}$, Y.~M.~Galperin$^{1,2}$, and T.~H.~Johansen$^{1,}$\cite{0}
}

\address{$^1$Department of Physics, University of Oslo, P. O. Box 1048 Blindern, 
0316 Oslo, Norway\\
$^2$A. F. Ioffe Physico-Technical Institute, Polytekhnicheskaya 26, 
St.Petersburg 194021, Russia}
\date{\today}

\maketitle

\begin{abstract}
Magnetization hysteresis loops and the ac susceptibility, $\chi = \chi' + {\rm i}\chi''$,
of a superconducting thin disk are calculated in the critical-state model
assuming a field-dependent critical current density, $J_c(B)$.
The results are obtained by solving numerically the set of coupled integral equations
for the flux and current distributions [PRB 60, 13112 (1999)] for a disk placed in a
perpendicular applied field $B_a$. From the magnetization curves
the range of fields where the vertical width of the loop, $\Delta M(B_a)$,
relates directly to $J_c(B_a)$ is determined.
The susceptibility is analyzed in the limits of small and large
ac-field amplitudes $\Bam$, and also as a parametric relation $\chi''(\chi')$.
Comparing our results with experimental data for $\chi''(\chi')$ shows that
by taking the $B$-dependence of $J_c$ into account the agreement improves dramatically,
in particular at small $|\chi'|$ (large field amplitudes).
We show that the asymptotic behavior for large $\Bam$ changes from $\chi' \propto \Bam^{-3/2}$ and
$\chi'' \propto \Bam^{-1}$ for the Bean model, to $\chi' \propto \Bam^{-3}$
and $\chi'' \propto \Bam^{-2}$ for $J_c$ decreasing with $|B|$ as
$|B|^{-1}$ or faster.
For small $\Bam$ the behavior can always be described by an effective Bean model with a
renormalized $J_c$.
We also find that in the $\chi''(\chi')$ plot
the peak of $\chi''$ increases in magnitude and shifts towards $\chi'=0$
when $J_c$ decreases with $|B|$.
This allows an easy experimental discrimination between a
Bean model behavior, one with $J_c(B)$, and
one where flux creep is an ingredient.

\end{abstract}


\narrowtext

\section{Introduction}

The critical state model (CSM) is widely accepted as a
powerful tool in the analysis of magnetic
properties of type-II superconductors.
In the parallel geometry, i.e., for long
samples like slabs and cylinders placed in a parallel magnetic field,
an extensive amount of theoretical work has already been carried
out. Exact results for flux density profiles,
magnetization,\cite{ChenM,Chad-dm,JohBra,Clem79} ac
susceptibility\cite{Clem79,Chen91,Forsthuber} etc., have been obtained for
a number of different field-dependent critical current densities.
During the last years even more attention has been paid
to the CSM analysis in the perpendicular geometry, i.e., for thin samples
in perpendicular magnetic fields.
Assuming a constant critical current (the Bean model),
explicit analytical results have been obtained for a
long thin strip~\cite{BrIn,Zeld} and a thin circular disk.\cite{Mik,Zhu,Clem,MikNote} 
{}From experiments, however, it is well known that also in such samples the critical current
density $j_c$ usually depends strongly on the local flux density $B$. 
Due to the lack of a proper theory, this dependence often hinders a precise interpretation of the 
measured quantities.\cite{Gomory,Stoppard,Herzog,Rao,Willemsen}

In the perpendicular geometry,
the ac susceptibility beyond the Bean model has been calculated only
by carrying out flux
creep simulations\cite{Br-ring,Br-disk2} assuming a power-law
current-voltage relation with a large exponent.  
However, quite recently an exact analytical approach was developed
for the CSM analysis of a long thin strip\cite{McD} and thin circular disk.\cite{we}
In both cases a set of coupled integral equations was derived for the flux and 
current distributions.
In the present paper we solve these equations numerically
for the thin disk case, and 
calculate magnetization hysteresis loops as well as the complex ac susceptibility. Results
for several commonly used $j_c(B)$ dependences are presented.  

The paper is organized as follows. 
In Sec.~\ref{be} we give a short description of the
exact solution for the disk problem. 
In Sec.~III, magnetization hysteresis loops are calculated and
the relation between the width of the loop and $j_c$ is discussed. 
The results for the complex ac susceptibility are presented in Sec.~IV and analysed with
emphasis on the asymptotic behavior at small and large field amplitudes. 
Finally, Sec.~V gives the conclusions. 

\section{Exact solution}
\label{be}

Consider a thin superconducting disk of radius $R$ and
thickness $d$, where $d \ll R$. 
We assume either that $d \ge \lambda$,
where $\lambda$ is the London penetration depth, or, if $d < \lambda$,
that $\lambda^2/d \ll R$. In the latter case the quantity $\lambda^2/d$ plays a
role of two-dimensional penetration depth.~\cite{DeGennes} We put the
origin of the coordinates at the disk center and direct the $z$-axis
perpendicularly to the disk plane.
The external magnetic field ${\bf B}_a$ is applied along the $z$-axis,
and the $z$-component of the field in the plane $z=0$ is denoted
as $B$. The current flows in the azimuthal direction, with a
sheet current denoted as $J(r)=\int_{-d/2}^{d/2} j(r,z)\, dz$, where
$j$ is the current density.

\subsection{Increasing field}

We begin with a situation where
the external field $B_a$ is applied to a zero-field-cooled disk.
The disk then consists of an inner  flux-free region, $ r \le a$ ,
and of an outer region, $ a < r \le R$, penetrated by magnetic flux.

{}In the CSM with a general $J_c(B)$
the current and flux density distributions in a disk are given by
the following coupled equations\cite{we}
\be
  J(r) = \left\{
  \begin{array}{lr}
  - \displaystyle{
    \frac {2r}{\pi} 
         \int_a^R\! \!
 dr' 
\sqrt{\frac{a^2 - r^2}{r'^2 -a^2}}\,\frac{J_c[B(r')]} {r'^2 - r^2}}
 , & r<a\\
 \nonumber\\
         - J_c[B(r)], &\! \! a<r<R
  \end{array}
  \right.
\label{JV}
\ee
\be
  B(r) = B_a + \frac{\mu_0 }{2\pi} \int^R_0 F(r,r') J(r')  dr'\,.
\label{BJ}
\ee
\be
   B_a = \frac{\mu_0}{2} \, \int_a^R \frac{dr}{\sqrt{r^2-a^2}} J_c[B(r)] \, .
\label{Ba}
\ee
Here
$F(r,r')\! = \!K(k)/(r  +  r')  -
E(k)/(r  -  r')$,  where $k(r,r') = 2\sqrt{rr'}/(r+r')$,
while $K$ and $E$ are complete elliptic integrals.
In the case of constant $J_c$, these equations reduce to the exact Bean-model formulas
derived in  Refs.~\onlinecite{Zhu} and~\onlinecite{Clem}.  
 
Note that the calculation  can be significantly simplified at
large external field where  $a \rightarrow 0$, and the critical state
$J(r)=J_c[B(r)]$ is established throughout the disk. The distribution
$B(r)$ is then determined by the single equation 
\begin{equation} \label{sm_a}
B(r)=B_a -\frac{\mu_0 }{2\pi} \int^R_0 F(r,r') J_c[B(r')]  dr'\, ,
\end{equation}
following from \eq{BJ}.

\subsection{Subsequent field descent}

If $B_a$ is
reduced, after being first raised to some maximum value $\Bam$, the flux
density will decrease in the outer part, $a \leq r \leq R$, and remain trapped 
in the inner part, see \f{f_scheme}.
We denote the flux front position, the current density and the
field distribution  at the maximum field as $a_m$, $J_m(r)$
and $B_m(r)$, respectively. 
Evidently, $J_m(r)$, $B_m(r)$, and $a_m$ satisfy
Eqs. (\ref{JV})-(\ref{Ba}).

Let the
field and current distributions during field descent be written as
\be
 B(r) = B_m(r) + {\tilde B}(r), \quad J(r) = J_m(r) + {\tilde J}(r).
\label{01}
\ee

The relation between ${\tilde B}(r)$ and ${\tilde
J}(r)$ then reads\cite{we}
\be
  {\tilde J}(r) = \left\{\!
  \begin{array}{lr}
  \displaystyle{
  \frac {2r}{\pi}
         \int_a^R \!dr' \ \sqrt{\frac{a^2 - r^2}{r'^2 - a^2}}\,
  \frac{\Jc2
  (r') } {r'^2 - r^2}} ,
         & r<a\\
         &\nonumber\\
         \Jc2(r), & a<r<R
  \end{array}
  \right.
\label{JR}
\ee
\be
  {\tilde B}(r) = B_a-\Bam + \frac{\mu_0 }{2\pi} \int^R_0 F(r,r') {\tilde J}(r')  
dr' \, .
\label{B1J}
\ee
\be
   B_a - \Bam = -  \frac{\mu_0}{2}\int_a^R\!  \frac{\Jc2 (r)}{\sqrt{r^2-a^2}} \, dr \, ,
\label{Ba1}
\ee
where we defined
\be
\Jc2(r) = J_c[B_m(r)+{\tilde B}(r)] + J_c[B_m(r)]\,. \label{Jeff}
\ee
 
Again, setting $J_c=$const, these equations reproduce the Bean-model
results.\cite{Zhu,Clem} 

If the field is decreased below $-\Bam$ the memory of 
the state at $\Bam$ is completely
erased, and the solution becomes equivalent to the virgin penetration case.
If the difference
$\Bam-B_a$ is large enough then one can again use 
\eq{sm_a}, only with the opposite sign in
front of the integral.  

Given the $J_c(B)$-dependence, a complete description
of any magnetic state 
is now found by solving the equations numerically. 
An efficient iteration procedure is described in Ref.~\onlinecite{we}.

\begin{figure} \vbox{
\centerline{ \psfig{figure=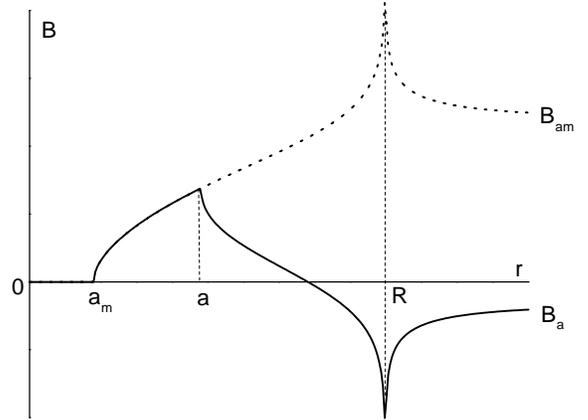,width=8cm}}
\caption{Flux density profile as the applied field
descends from a maximum value $\Bam$.
\label{f_scheme}}}
\end{figure}

\section{Magnetization}
\label{Magnetization}

The magnetization of a disk is defined as
the magnetic moment, $\pi \int_0^R\! r^2\, J(r)\, dr$, per unit volume. Due to
symmetry the magnetization is directed along the $z$-axis. In
a fully penetrated state described by the Bean model with critical current 
$J_{c0}$, the magnetization equals
$M_0=J_{c0}R/3d$. It is convenient to use $M_0$ for normalization,  i.e.
\be \label{nm}
  \frac{M}{ M_0} = \frac{3}{R^3} \int_0^R \! \frac{J(r)}{J_{c0}}\,r^2\, dr.
\ee
The magnetization can be calculated using the current profiles
obtained by the procedure described in the previous section.
Shown in \f{f_loops} are magnetization hysteresis loops calculated
for the $J_c(B)$-dependences:

\begin{figure} \vbox{
\centerline{\psfig{figure=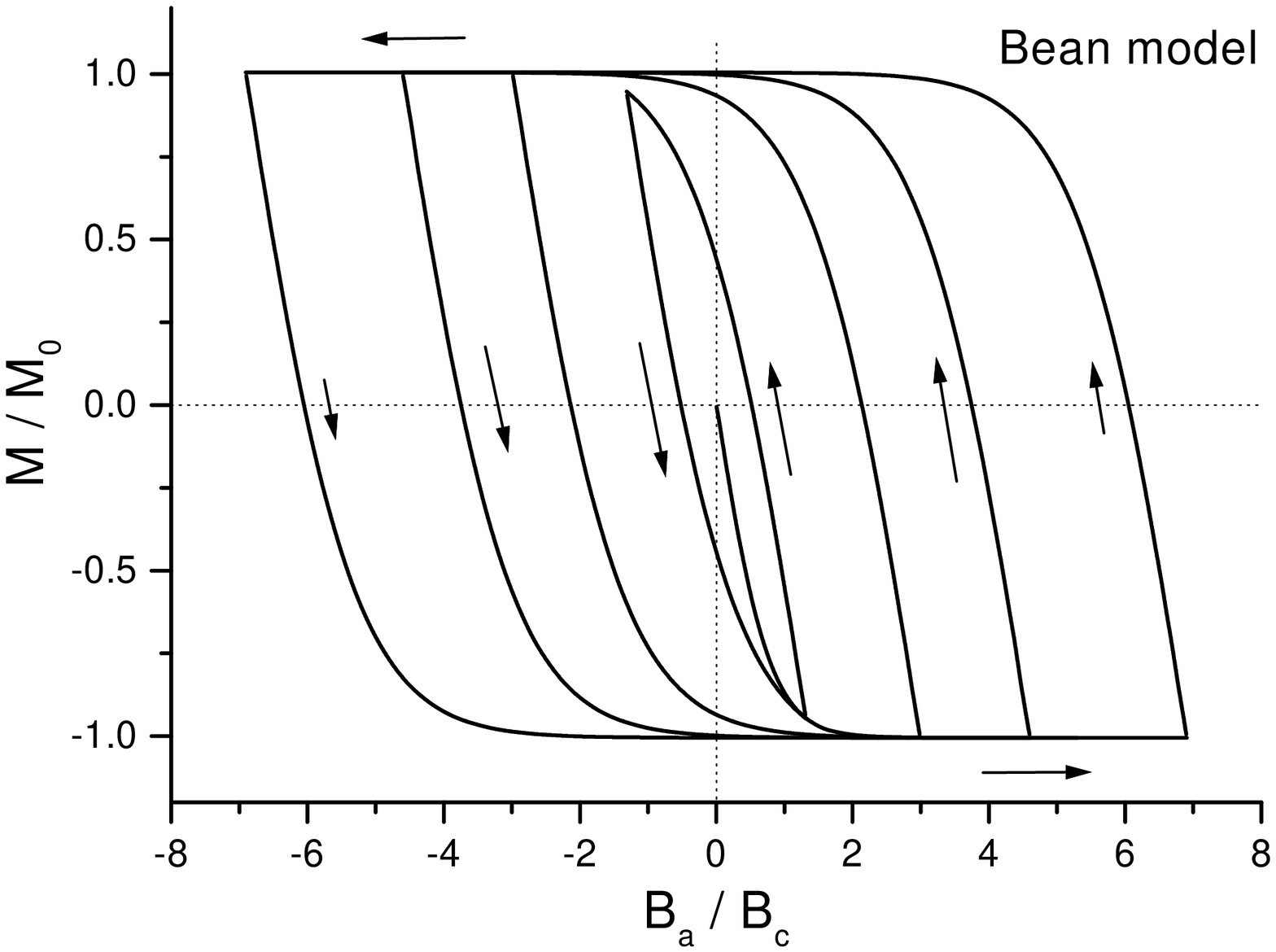,width=8cm}}
\centerline{\psfig{figure=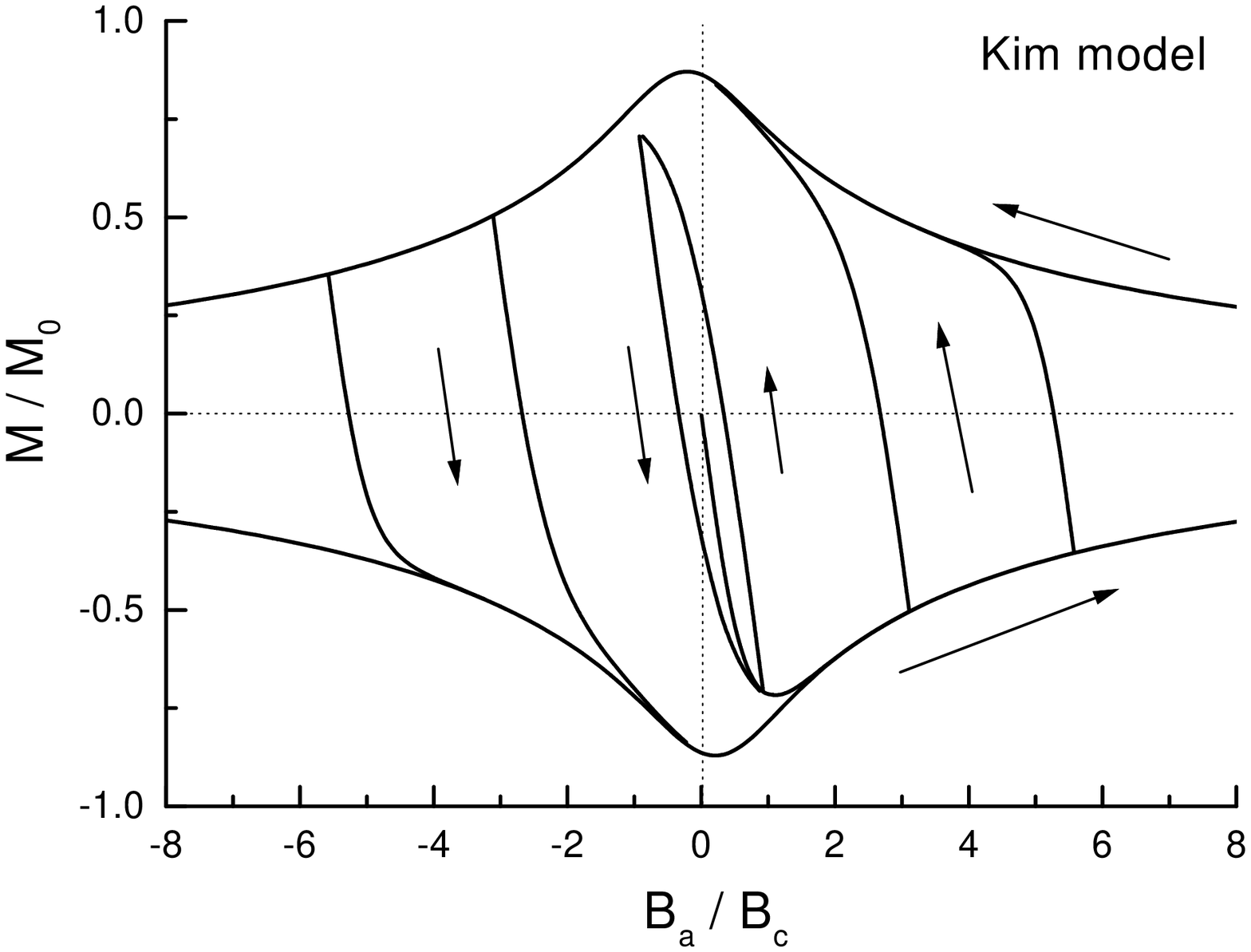,width=8cm}}
\centerline{\psfig{figure=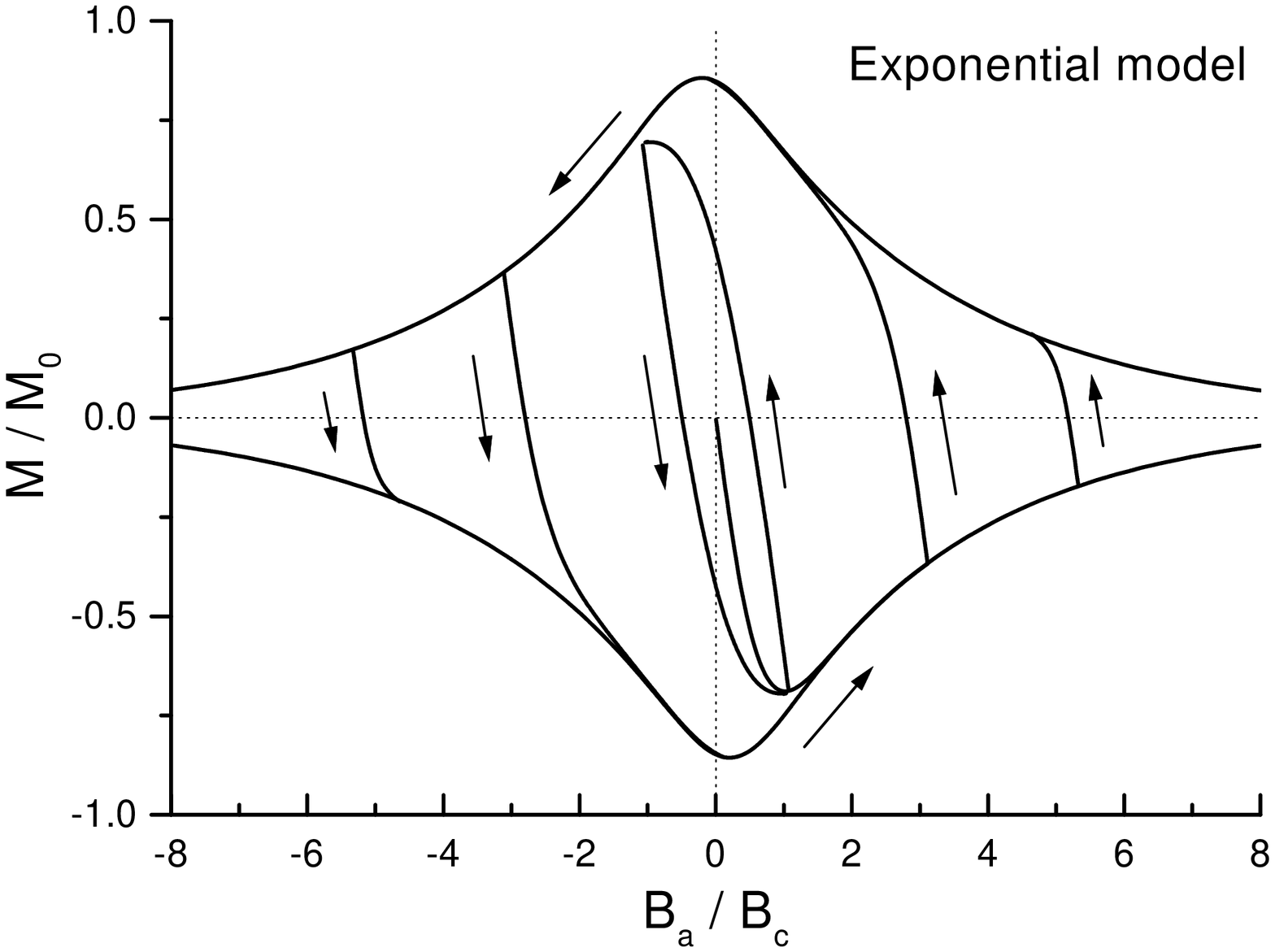,width=8cm}}
\caption{Magnetization hysteresis loops for a thin disk
for the Bean model ($J_c = const$), the
Kim model, \eq{km},  and the exponential model, \eq{em}, the last two both
with $\Bn=3B_c$. The parameters $B_c$ and $M_0$ are defined in the text.
\label{f_loops}}}
\end{figure}

\bea
J_c&=& J_{c0}/(1+|B|/\Bn) \quad \quad \ \text{(Kim model),} \label{km} \\
J_c&=& J_{c0}\, \exp(- |B|/\Bn) \qquad   \text{(exponential model).}
\label{em}
\eea

A striking manifestation of the $B$-dependence is a peak occuring at small
$B_a$. The calculations show
that for any choice of the parameter $B_0$, the peak is always
located at {\em negative\/} $B_a$ on the
descending branch of the major loop. Such a peak position at negative $B_a$ is
a typical feature also in the parallel geometry.\cite{ChenM,Chad-dm,JohBra}
However, it contrasts the case of a
thin strip in perpendicular field, where it was shown analytically\cite{prl} that
for any $J_c(B)$-dependence the peak is located {\em exactly\/} at $B_a=0$.

\begin{figure}[b] \vbox{
\centerline{\psfig{figure=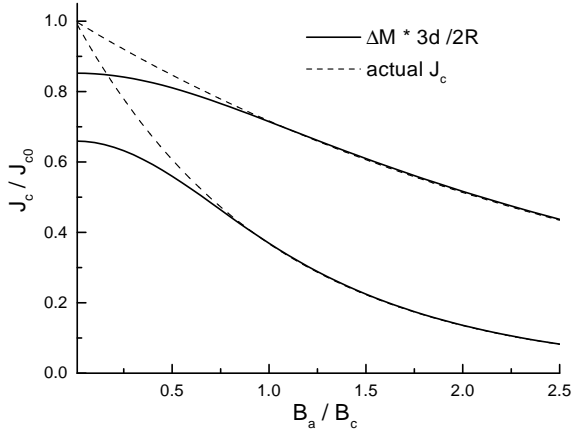,width=8cm}}
\caption{The critical current inferred from the width, $\Delta M$, of the
major magnetization loop using \eq{dm}. For comparison the plot also shows
the actual $J_c(B)$ used in the calculation: the exponential model with $B_0=B_c$
(lower curves) and $B_0=3B_c$ (upper curves).
The agreement is excellent for fields larger than $B_c$.
\label{f_dm}}}
\end{figure}

In the Bean model, there is a simple relation between the
critical current and the width $\Delta M$ of the major magnetization loop,
\be
J_c = \frac{3d}{2R} \Delta M .
\label{dm}
\ee
The same expression is often used to determine $J_c$
from experimental $\Delta M$ data even when the width of the observed loop is not constant.
As discussed in Refs.~\onlinecite{ChenM,Chad-dm,JohBra}
the applicability range of such a procedure is limited. In the parallel geometry
a simple proportionality only applies for $B_a$ larger than
the full penetration field.
For the thin disk case the field
range where $J_c \propto \Delta M$ can be estimated from our calculations.
Figure~\ref{f_dm} shows $J_c(B)$ inferred from the magnetization loop using \eq{dm},
together with the actual $J_c(B)$.
One can see that at fields larger than the characteristic field
\be
B_c \equiv \mu_0 J_{c0}/2 \ ,
\label{Bc}
\ee
there is essentially
no distinction between the two curves. We find that this holds
independently of $B_0$ and also for other $J_c(B)$ models.
Therefore, also for the present geometry the $B$-dependence
of $J_c$ can be inferred directly from $\Delta M(B_a)$,
except in the low-field region. Here the correct $J_c(B)$ can be
obtained only by a global fit of the magnetization curve.

The Bean-model virgin magnetization for a thin disk can be expanded in $B_a$ as\cite{Clem,Br-disk2}
\be
-\mu_0 M \approx \chi_0 B_a \l( 1 - \frac 12 \l( \frac{B_a}{B_c}\r)^2 \r) ,
\label{MBean}
\ee
where $\chi_0 = 8R/3\pi d$ is the Meissner state susceptibility.
Our numerical calculations show that
the same expansion also holds for $B$-dependent $J_c$, only with an effective value $\Bceff$
satisfying
\be
\label{if}
\Bceff / B_c \approx 1-\alpha \, \sqrt{ B_c / \Bn } .
\ee
We find that if $\Bn/B_c \ge 0.5$ the parameter
$\alpha = 0.50$ for the exponential model, and $\alpha = 0.43$
for the Kim model. In the parallel geometry
the low-field expansion has an additional $B_a^2$ term which is not affected
by the $J_c(B)$ dependence.\cite{JohBra} Thus, the deviation from the Meissner response
at small $B_a$ is there insensitive to $J_c(B)$.
This result contrasts the case of perpendicular geometry
where due to demagnetization effects, a $B$-dependence of $J_c$
affects the flux behavior even in the limit of low fields, see discussion
in Ref.~\onlinecite{we}.

\section{Complex ac susceptibility}
\label{loss}

\subsection{Basic expressions}

The hysteretic dependence of the magnetization $M$ as the
applied field $B_a$ is cycled
leads to ac losses. The energy dissipation per cycle of $B_a$
is
\be \label{wv}
W=\int_{\rm cycle} B_a(t) \frac{d M (t)}{dt} \, dt \,
\ee
per unit volume.
According to the critical state model, $M(t)$ follows $B_a(t)$
adiabatically, i.e., $M(t)=M[B_a(t)]$. Thus, the losses are given by the area of the
magnetization hysteresis loop, $\oint M\, dB_a$.

It is conventional to express the ac response through
the imaginary and real parts of the so-called nonlinear magnetic
susceptibility.\cite{Gomory}
If the applied field is oscillated harmonically with amplitude
$\Bam$, i.e., $B_a (t)=\Bam\, \cos \omega t$, the
magnetization  is also oscillating with the same period.
The complex susceptibility is then defined by the coefficients of the
Fourier series of the in general anharmonic $M(t)$,
where the real and imaginary parts are given by
\begin{eqnarray*}
\chi'_n&=&\frac{\mu_0 \omega}{\pi \Bam}\int_0^{2\pi/\omega} \! M(t)
\cos (n \omega t) \, dt \ , \\
\chi''_n&=&\frac{\mu_0 \omega}{\pi \Bam}\int_0^{2\pi/\omega} \! M(t)
\sin (n \omega t) \, dt \ ,
\end{eqnarray*}
respectively.

The dissipated energy, $W$, is determined by the response $\chi''_n$
at the fundamental frequency, namely
\be \label{chi''}
\chi''\equiv \chi''_1=\frac{\mu_0 W}{\pi \Bam^2}= \frac{2\mu_0}{\pi
\Bam^2}\int_{-\Bam}^{\Bam}\! M(B_a)\, dB_a\, . 
\ee
Below we shall also analyze the real part of the susceptibility at the
fundamental frequency, $\chi'\equiv \chi'_1$, which
can be expressed as
\be \label{chi'}
\chi' =-\frac{2\mu_0}{\pi \Bam^2}\int_{-\Bam}^{\Bam}\! \frac{M(B_a)\, B_a\,
dB_a}{\sqrt{\Bam^2-B_a^2}}\, . 
\ee
The $\chi''(\Bam)$ and $\chi'(\Bam)$  are calculated from these expressions 
using $M(B_a)$ 
obtained by the previously described procedure with $\Bam$ covering a wide range of amplitudes. 
For convenience, we normalize the susceptibilities  to the Meissner state 
value $\chi_0=8R/3\pi d$.~\cite{Clem}  

\begin{figure}\vbox{
\centerline{\psfig{figure=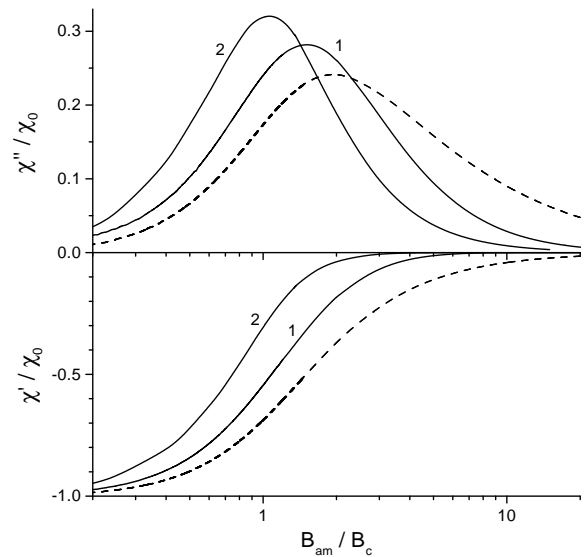,width=8cm}}
\caption{\label{f_chi}  Real (bottom) and imaginary (top) parts of the nonlinear
susceptibility for a thin disk as functions of the amplitude $\Bam$ of the applied ac
field. Calculations are based on \eqs{chi''}{chi'} with $J_c(B)$  
given by the exponential model, \eq{em}, with $\Bn=3B_c$ (curve~1) and
$\Bn=B_c$ (curve~2). For comparison the results for the Bean model are
also shown (dashed line).}}
\end{figure}

As seen from \f{f_chi}, the response $\chi''$ shows a maximum
as a function of the field amplitude. 
Such a maximum is in fact a common feature in all geometries.
For the Bean model for a long cylinder the peak is
known to occur when $\Bam$ is equal to the full penetration field.
In the perpendicular geometry the interpretation of the peak position is
not so simple. Even in the Bean model for a thin disk only numerical results
are available\cite{Clem}: the peak value equals
$\chi''_{\max}=0.24$ and occurs at an amplitude of $\Bam=1.94 B_c$,
corresponding to the penetration $1-a_m/R=72\%$.
We find that the $B$-dependence of $J_c$ leads to a slight increase
both in $a_m$ and in the peak magnitude.
For example, the numerical results for the Kim model with $B_c = B_0$
give $\chi''_{\max}=0.29$ and $1-a_m/R=70\%$.
The difference between various $J_c(B)$ models becomes more distinct
if one analyses the asymptotic behavior at small and large field amplitudes,
as shown below.

\subsection{Low-field behavior}

At small field amplitudes the Bean model gives the exact expressions\cite{Clem,Br-disk2}
\bea
\label{chi1_as}
\chi' /\chi_0 &=& -  1 + 15 (\Bam/B_c)^2 /32,  \\
\chi'' / \chi_0&=&  (\Bam/B_c)^2 / \pi \ .
\label{chi_as}
\eea
Shown in  \f{f_low2} are our numerical results for $\chi''$ for
the exponential model. From the log-log plot it is clear that the
quadratic dependence on $\Bam$ retains as in \eq{chi_as} only
with a modified coefficient.
Moreover, we find that also $\chi'$ can be described
by the Bean model expression \eq{chi1_as} with the {\em same} effective $B_c$.
The effective $B_c$ fits the expression
\be \label{if2}
{\Bceff}/{B_c} = 1-\alpha \, { B_c}/{  \Bn} \, ,
\ee
when $B_0/B_c \ge 1$ with $\alpha = 0.42$ for the exponential model, and $\alpha = 0.36$
for the Kim model. Interestingly, the same effective description was found
for the flux penetration depth $a$,\cite{we} whereas it deviates from the description of the virgin magnetization, \eq{if}.

\begin{figure}\vbox{
\centerline{\psfig{figure=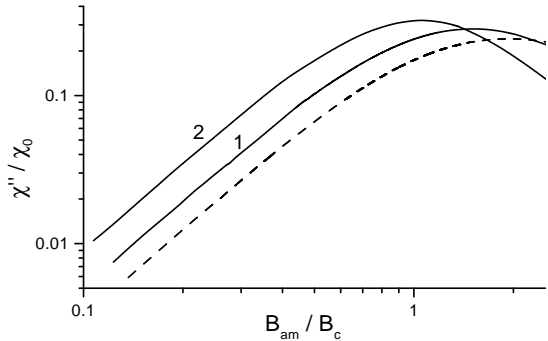,width=8cm}}
\caption{Imaginary part of the
susceptibility for the exponential model with
$\Bn/B_c=3$  (curve~1) and with $\Bn/B_c=1$ (curve~2).
At low fields both curves as well as the dashed curve presenting the
Bean model result follow the same quadratic law,
$\chi'' \propto \Bam^2$.\label{f_low2}}}
\end{figure}

\subsection{High-field behavior}

The high-field behavior of the dissipated energy $W$ is shown in
\f{f_high2} for a variety of $J_c(B)$ dependences.
We choose to plot $W$ rather than $\chi''$ because the
difference between the asymptotic behavior in the
various models becomes more evident.
One sees from the figure that for large $\Bam$ the Bean model yields $W \propto \Bam$.
The exponential model 
shows saturation, whereas one finds after a closer inspection that the Kim model
leads to a logarithmic increase. 
These behaviors can be understood by considering the fact that 
for large amplitudes the disk is fully
penetrated and $B(r)\approx B_a$. Therefore, $M(B_a) \propto J_c(B_a)$,  
and one obtains $W \propto \int^{\Bam} J_c(B) \, dB$.

\begin{figure}\vbox{
\centerline{\psfig{figure=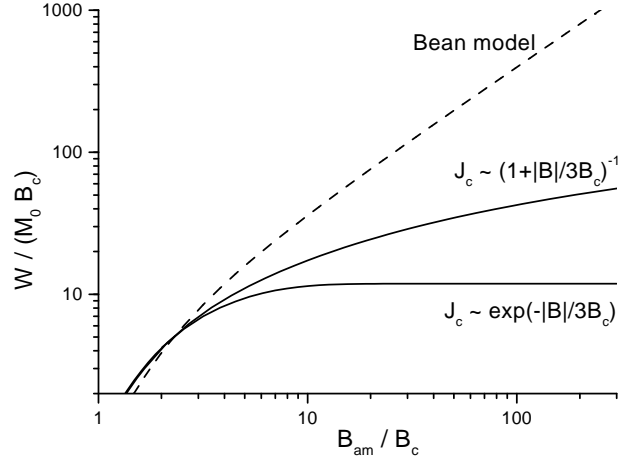,width=8.6cm}}
\caption{ High-field behavior of the dissipated energy,
$W$, for the Bean, Kim and exponential models.\label{f_high2}}}
\end{figure}

The high-field behavior of the real part of the susceptibility, $\chi'$,
for different $J_c(B)$ is shown in \f{f_high1}.
For the Bean model we find asymptotically that
$\chi'/\chi_0 = - 1.33 (\Bam/B_c)^{-3/2}$ (dotted line),
which is in agreement with Eq.~32 in Ref.~\onlinecite{Clem}.
For the $B$-dependent $J_c$'s we also find power-law behavior, although
with different exponents. For both the Kim and exponential model
the asymptotic behavior is described by $\chi' \propto \Bam^{-3}$.
However, also intermediate values for the exponent are possible, e.g.,
for $J_c=J_{c0}/[1+(|B|/3B_c)^{1/2}]$ the numerical results
suggest that $\chi' \propto \Bam^{-9/4}$.

In order to understand this power-law behavior let us rewrite \eq{chi'} as
\be
\chi' =
\frac{2}{\pi \Bam^2}\int_{0}^{\Bam}\! \frac{\mu_0 \Mrev (B_a)\,
B_a\, dB_a}{\sqrt{\Bam^2-B_a^2}}\, ,
\ee
where $\Mrev = M_{\uparrow} + M_{\downarrow}$ is the reversible magnetization.
The integrand has different estimates in the
regions~I, II, and~III indicated in \f{f_range}. Therefore we divide the interval of
integration correspondingly, $\chi' = \chi'_{I} + \chi'_{II} + \chi'_{III}$.
In region~I,  $\Mrev$ does not depend
on $\Bam$, thus, $\chi'_I \propto \Bam^{-3}$ at large $\Bam$.
In region~II ($B_a \gg B_c$) we use that
$$\Mrev(B_a) \propto \int \! dr\, r^2 \left[ J_c(B_a+B_i(r))-J_c(B_a-B_i(r))\right],$$
where  $B_i$ is the field created by the
current. Expanding this expression one has $\Mrev \propto J_c'(B_a) \int \! dr\, r^2 B_i(r)$.
Then using the further simplification that $\int \! dr\, r^2 B_i(r) \propto J_c(B_a)$,
one obtains
$$
\chi''_{II} \propto \frac{1}{\Bam^2} \int_{II} \!\frac{J_c(B_a) \, J_c'(B_a) \ B_a}
{\sqrt{\Bam^2-B_a^2}}\, dB_a
\, .
$$ 
Taking $J_c(B)\propto (\Bn/B)^{s}$ at large $B$, we
arrive at the estimates, $\chi'_{II} \propto \l(B_c/\Bam \r)^2  (\Bn/\Bam)^{2s}$ for small $s$,
and $\chi'_{II} \propto \l(B_c/\Bam \r)^3  (\Bn/B_c)^{2s}$ for large $s$.
Finally, consider the region III, where  $\Bam - B_a$ is of the order of~$\mu_0 J_c(B_a)$.
Since the initial slope of the return branch
does not depend on $\Bam$, we have that $\Mrev(B_a) \propto \Bam-B_a$.
It then follows that, $\chi'_{III}  \propto \l(B_c/\Bam \r)^{3/2}
\left(\Bn/\Bam\right)^{3s/2}$.
As the asymptotic behavior at large $\Bam$ is determined by the
slowest decaying term, we arrive at the following result
\be
  \chi' \propto \left\{
  \begin{array}{lr}
  \displaystyle{ \Bam^{-3(1+s)/2}} , & s<1\\
&  \nonumber \\
         \Bam^{-3}, & \! s \ge 1
  \end{array}
  \right.
\label{chi1as}
\ee

\begin{figure}\vbox{
\centerline{\psfig{figure=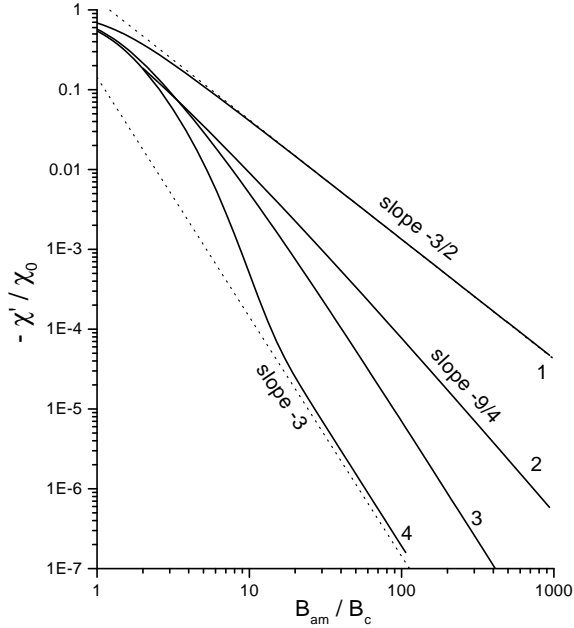,width=8cm}}
\caption{ High-field behavior of the real part of the
susceptibility, $\chi'$, for the Bean model (curve 1) and for different
$J_c(B)$ dependences;
$J_c=J_{c0}/(1+(|B|/3B_c)^{1/2})$ (curve 2); Kim model (curve 3)
and exponential model (curve 4), both with $\Bn=3B_c$.\label{f_high1}}}
\end{figure}

These power-laws fully agree with our numerical calculations shown in \f{f_high1}.
The expression \ref{chi1as} gives the exact values
for the exponent found for the Bean model ($s=0$),
the Kim ($s=1$) and exponential ($s=\infty$) models and even
for the $J_c(B)$ with $s=1/2$. Note however, that
this asymptotic behavior is sometimes established only at rather low
values of $|\chi'|$, see curve~4 in Fig.~\ref{f_high1}.
Therefore one should be very careful in interpretation of
corresponding experimental log-log plots.

It should be specially emphasized that the presented analysis for the
high-field asymptotic behavior is not restricted to a thin disk.
In fact, we expect the result (\ref{chi1as})
to be valid in {\em any geometry}.
This result is also in agreement with numerical calculations
for long samples described by the Kim and the exponential model.\cite{Chen91}
\begin{figure} \vbox{
\centerline{\psfig{figure=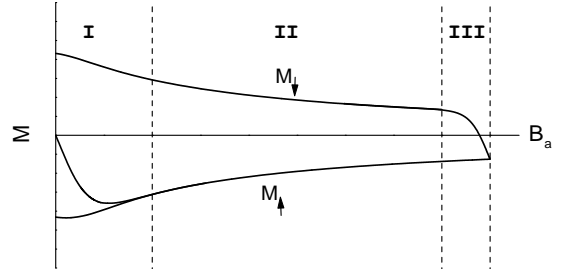,width=8cm}}
\caption{Division of a magnetization loop into 3 regions treated differently
when estimating $\chi'$ at large $\Bam$. \label{f_range}}}
\end{figure}

\subsection{Plots of $\chi''$ versus $\chi'$}

\begin{figure}\vbox{
\centerline{\psfig{figure=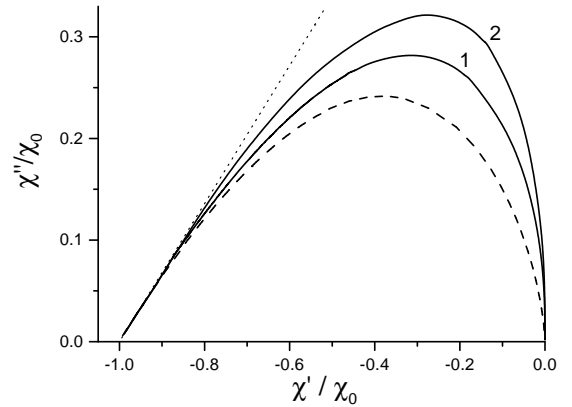,width=8cm}}
\caption{\label{f_chichi} Parametric plot of the 
complex susceptibility for a thin disk using the
three sets of curves presented in \f{f_chi}. The dotted line
shows the Bean model asymptotic behavior, \eq{uni}.
A $B$-dependence of $J_c$ substantially distorts the plot,
although not the slope at $\chi' \rightarrow -1$.}}
\end{figure}

In contrast to graphs of $\chi$ as a function of the field amplitude
or temperature,
a plot of $\chi''$ versus $\chi'$ contains only dimensionless quantities,
and is therefore
very useful for analyzing experimental data\cite{Stoppard,Herzog}. In practice,
such a parametric plot $\chi''(\chi')$  can be obtained
by scans either over the magnetic field amplitude or over the temperature.
Figure~\ref{f_chichi} presents the
$\chi''(\chi')$ plot of the data shown in \f{f_chi}.
We observe that a $B$-dependence of $J_c$
gives a significant distortion of the graph. Compared
to the Bean model one finds that: (i) the maximum is shifted
to higher values of $\chi''$; (ii) it occurs
at smaller values of $-\chi'$; (iii) in the limit of large $\Bam$
(or high temperatures)
$\chi''$ falls to zero more abruptly.

Meanwhile, at small $\Bam$, as $\chi' \rightarrow -1$,
the slope of $\chi''(\chi')$ curve remains the same as
in the Bean model, namely,
\be
\frac{\chi''}{\chi_0} = \frac{32}{15\pi}
\l(1+ \frac{\chi'}{\chi_0} \r) \ \quad {\rm at} \quad \chi' \rightarrow -1 \ .
\label{uni}
\ee
This result holds for {\em any} $J_c(B)$. It also follows
from the previous analysis showing that at low fields both $\chi'$ and $\chi''$ are
modified by $J_c(B)$ in the same way.
The universal slope given by \eq{uni} allows one to examine if
experimental data are described by the critical state model
without {\em a priori} knowledge of the actual $J_c(B)$ dependence for the sample.

The presented $\chi''(\chi')$ plots for a disk in a perpendicular field
should be compared to similar plots
for the long samples in a parallel field studied systematically in
Ref.~\onlinecite{Chen91}.
As expected, the Bean-model curve for a thin disk shown by
the dashed line in our Fig. 9 appears quite
different
from the Bean-model curves for long samples shown in Fig.~7(a,b) of
Ref.~\onlinecite{Chen91}.
Meanwhile, further analysis of these figures shows that
the account of a $B$-dependent $J_c$ always leads to very
similar distortions of the $\chi''(\chi')$ plots.
Namely, in all geometries the $\chi''$ peak
increases in magnitude and shifts towards $\chi'=0$.
Note that such a behavior is found when
the characteristic field $B_0$ of
the $J_c(B)$-dependence is larger or of the order of $B_c$.
For $B_0 \ll B_c$ this behavior may change qualitatively.
In particular, in the parallel geometry, the peak position,
$\chi'_{\max}$, becomes a nonmonotonous function
of $B_0$.\cite{Chen91}
However, the case of $B_0 \ll B_c$ is not very realistic
for a thin disk since $B_c$ is proportional to the
sample thickness while $B_0$ is usually taken as geometry-independent.

\begin{figure}\vbox{
\centerline{\psfig{figure=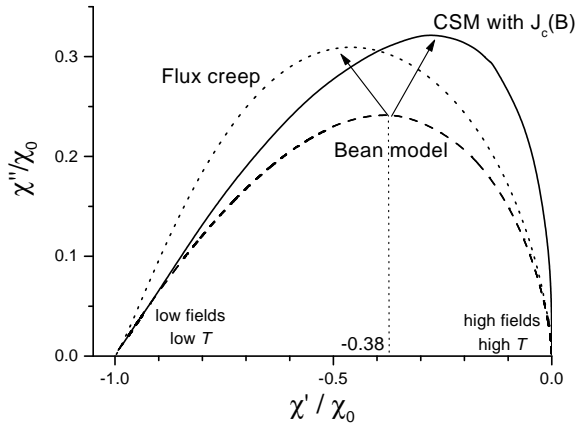,width=8cm}}
\caption{\label{f_final} The behavior of $\chi''(\chi')$ for various models.
The Bean model predicts a peak located at $\chi'_{\max}=-0.38$.
A $B$-dependence in $J_c$ shifts the peak to the right and
changes the behavior at $\chi' \rightarrow 0$  (our results),
while flux creep shifts it to the left and changes
the behavior at  $\chi' \rightarrow -1$
(Ref.~\p\onlinecite{Br-disk2}).  }}
\end{figure}

It is interesting to compare our $\chi''(\chi')$ plots to the ones obtained
by calculations based on a non-linear
current-voltage curve, $j \propto E^{1/n}$, $n<\infty$.
Shown in  \f{f_final} together with the CSM results is
a $\chi''(\chi')$-curve (dotted line)
drawn in accordance to  typical graphs presented in
Refs.~\onlinecite{Br-ring,Br-disk2}.
Compared to the Bean model curve,  the maximum of $\chi''$
increases in magnitude and shifts towards $\chi'= -1$.
Moreover, the slope at $\chi' \rightarrow -1$ becomes steeper.
The last two features are in a strong contrast to the effect of having a $B$-dependent $J_c$
in the CSM.
Consequently, an analysis of the $\chi''(\chi')$ plot allows one
to discriminate between a strict CSM behavior and one where flux creep
is an ingredient.

\begin{figure}\vbox{
\centerline{\psfig{figure=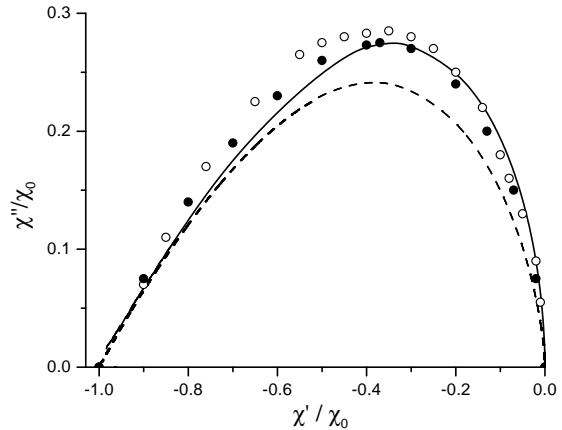,width=8cm}}
\caption{\label{f_fitexper}
Experimental susceptibility data from Ref.~\p\onlinecite{Stoppard}
(open circles) and Ref.~\p\onlinecite{Herzog} (solid circles)\p\cite{strange}
together with the CSM predictions for a thin disk: the Bean model (dashed curve) and the
Kim model with $\Bn=3B_c$ (full line). The Kim model gives a better
agreement with experiment over the whole range.}}
\end{figure}

Finally, we compare in Fig.~\ref{f_fitexper} our theoretical results to
available experimental data on the susceptibility of
YBaCuO films.\cite{Stoppard,Herzog,strange} The shown data  were obtained
by reading selected points in the graphs found in the literature.
It is evident that the poor fit by the Bean model (dashed curve) is greatly improved
by the curve (full line) calculated for a $B$-dependent $J_c$.  Whereas the agreement is
better throughout the $\chi''(\chi')$ plot, it
is especially evident at small $|\chi'|$ (large field amplitudes), where
the $J_c(B)$-dependence plays a major role.
There is still a discrepancy in the low-field region, where
all experimental points do not follow the universal CSM slope given by
Eq.~\ref{uni}.
The deviation can be caused by a flux creep leading to a steeper
slope.\cite{Br-ring,Br-disk2}
This suggestion can be checked experimentally by analyzing
$\chi''(\chi')$ plots obtained at different temperatures.

\section{Conclusion}

Magnetization and ac susceptibility of a thin
superconducting disk placed in a perpendicular magnetic field
were analyzed in the framework of the critical
state model where $J_c$ depends on the local flux density.
We solved numerically the set of coupled integral
equations for the flux and current distributions,
and from that calculated magnetization hysteresis loops as well as the
susceptibility, $\chi = \chi' + {\rm i}\chi''$.
The results, which were obtained for several commonly used $J_c$ decreasing with $|B|$,
allowed us to determine the range of fields where the vertical width of the
major magnetization loop, $\Delta M(B_a)$, is directly related to $J_c(B_a)$. 

We have shown that at small fields the virgin magnetization and complex susceptibility
have the same dependence
on $B_a$ as for the Bean model, although with different coefficients.
For large ac amplitudes, $\Bam$, the behavior of the ac susceptibility
changes from $\chi' \propto \Bam^{-3/2}$ and
$\chi'' \propto \Bam^{-1}$ for the Bean model, to $\chi' \propto \Bam^{-3}$
and $\chi'' \propto \Bam^{-2}$ for $J_c$ decreasing with $|B|$ as
$|B|^{-1}$ or faster.
We could show numerically, and also
presented an argument, that when asymptotically
$J_c \sim |B|^{-s}\, , \, s<1$,
one has  $\chi' \propto  \Bam^{-3(1+s)/2}$.
The results for the high-field behavior of the susceptibility are
expected to be valid for superconductors of {\em any geometry}.

A most convenient test for critical-state models is provided by an
analysis of the $\chi''(\chi')$ plot.
We conclude that the asymptotic behavior at $\chi' \rightarrow -1$
is universal for the CSM with {\em any} $J_c(B)$. On the other hand,
flux creep can affect this behavior. The peak in $\chi''$ at $\chi'=-0.38$
predicted by the Bean model
was found to be shifted toward $\chi'=0$ due to the $B-$dependence in $J_c$,
and toward $\chi'=-1$ because of flux creep.

\acknowledgements

The financial support from the Research Council of Norway (NFR), and from
NATO via NFR is gratefully acknowledged.


\widetext 
\begin{references}

\bibitem[*]{0}Email: t.h.johansen@fys.uio.no

\bibitem{ChenM} D. X. Chen and R. B. Goldfarb,
J. Appl. Phys. {\bf 66}, 2489 (1989);
D.~X.~Chen, A.~Sanchez, J.~S.~Minoz, J. Appl. Phys. {\bf 67}, 3430 (1990).

\bibitem{Chad-dm} P. Chaddah, K. V. Bhagwat, and G. Ravikumar, Physica C {\bf 159}, 570 (1989).


\bibitem{JohBra} T. H. Johansen and H. Bratsberg,
J. Appl. Phys. {\bf 77}, 3945 (1995).

\bibitem{Clem79} J. R. Clem, J. Appl. Phys. {\bf 50}, 3518 (1979).

\bibitem{Chen91} D. X. Chen and A. Sanchez, J. Appl. Phys. {\bf 70}, 5463 (1991).

\bibitem{Forsthuber} M. Forsthuber and G. Hilscher, Phys. Rev. B
{\bf 45}, 7996 (1992).

\bibitem{BrIn}  E. H. Brandt, and M. Indenbom, Phys. Rev. B {\bf 48}, 12893
(1993).

\bibitem{Zeld} E. Zeldov,  J. R. Clem, M. McElfresh,
and M. Darwin,  Phys. Rev. B {\bf 49}, 9802 (1994).

\bibitem{Mik} P. N. Mikheenko and Yu. E. Kuzovlev,
Physica C {\bf 204}, 229 (1993).

\bibitem{Zhu} J. Zhu, J. Mester, J. Lockhart, and J. Turneaure,
Physica C {\bf 212}, 216 (1993).

\bibitem{Clem} J. R. Clem and A. Sanchez, Phys. Rev. B {\bf 50},
9355 (1994).

\bibitem{MikNote} Some final expressions obtained in Ref.~\onlinecite{Mik}
are not correct, see discussion in Ref.~\onlinecite{Zhu}.

\bibitem{Gomory} F. G\"om\"ory, Supercond. Sci. Technol. {\bf 10}, 523 (1997).

\bibitem{Stoppard} O. Stoppard and D. Gugan, Physica C {\bf 241}, 375 (1995).

\bibitem{Herzog} Th. Herzog, H. A. Radovan, P. Ziemann, and E.~H.~Brandt,
Phys. Rev. B {\bf 56}, 2871 (1997).

\bibitem{Rao} B. J. J\"onsson, K. V. Rao, S. H. Yun, and
U. O. Karlsson, \prb {\bf 58}, 5862 (1998).

\bibitem{Willemsen} B. A. Willemsen, J. S. Derov, and S.~Sridhar,
Phys. Rev. B {\bf 56}, 11989 (1997).

\bibitem{Br-ring} E. H. Brandt, \prb {\bf 55}, 14513 (1997).
\bibitem{Br-disk2} E. H. Brandt, \prb {\bf 58}, 6523 (1998).

\bibitem{McD} J. McDonald and J. R. Clem, Phys. Rev. B {\bf 53}, 8643 (1996).

\bibitem{we} D. V. Shantsev, Y. M. Galperin, T. H. Johansen, \prb {\bf 60}, 13112 (1999).

\bibitem{DeGennes} P. G. de Gennes, {\it Superconductivity of Metals
and Alloys\/} (Benjamin, New York, 1966).

\bibitem{prl} D. V. Shantsev, M. R. Koblischka, Y. M. Galperin,
T. H. Johansen, L. Pust, and M. Jirsa, \prl {\bf 82}, 2947 (1999).

\bibitem{strange}
The Bean-model $\chi''(\chi')$-curves for all geometries
presented in Fig.~2 of Ref.~\p\onlinecite{Stoppard}
and Fig.~4 of Ref.~\p\onlinecite{Herzog} are incorrect.
At large field amplitudes , i.e., at $\chi' \rightarrow 0$,
they show a linear behavior, while the true
Bean-model behavior is always $\chi'' \propto |\chi'|^{2/3}$
(see e.g. discussion in Ref.~\p\onlinecite{Br-disk2} or
exact results for the disk case in Eqs.~(32) and~(33) of
Ref.~\p\onlinecite{Clem}).


\end{references}
\end{document}